# SINAP surface preparation processing for 500MHz superconducting cavity

Zhenyu Ma[1,3], Haibo Yu[1,2,3], Dongqing Mao[1,3], Hongtao Hou[1,2,3], Ziqiang Feng[1,3], Chen Luo[1,3], Shenjie Zhao[1,3], Yubin Zhao[1,3], Zhigang Zhang[1,2,3], Xiang Zheng[1,3], Zheng Li[1,3], Bo Yin[1,2,3], Jianfei Liu[1,3,*]

1 (Shanghai Institute of Applied Physics, CAS, Shanghai 201800, China)
2 (Graduate University of Chinese Academy of Sciences, Beijing 100049, China)
3 (Shanghai Key Lab of Cryogenics & Superconducting RF Technology, Shanghai 201800, China)

**Abstract**
This paper illustrates the design, fabrication and experiment results of surface preparation system for 500MHz superconducting cavity at Shanghai Institute of Applied Physic (SINAP). The SINAP established a set of clean room, buffered chemical polishing equipment, and high pressure ultra-pure water rinsing facility. The whole surface preparation procedure has been operated successfully and verified by the successful vertical tests of 500MHz single cell superconducting cavity.

## 1 INTRODUCTION

Superconducting (SC) niobium cavity is one of the major components of a particle accelerator. It is used for accelerating charged particles to compensate for beam loss or to hit the target material. Comparing with normal conducting copper cavities, very high accelerating field gradient of the order of 50MV/m with small power dissipation can be achieved with SC cavities[1]. Furthermore, SC niobium cavity has a very high $Q$-value, therefore its high gradient performance is sensitive to the surface imperfections: damaged layer, scratches, inclusions, electron beam welding defects, and contaminations, which can arose cavity loss mechanisms such as field emission or thermal breakdown[2]. The role of surface preparation with SC cavity is to eliminate such problems. There exist several methods of surface preparation of niobium cavities, such as barrel polishing[3], buffered chemical polishing (BCP)[4,5], Electropolishing[6], high pressure rinsing (HPR)[7,8], low or high temperature baking[9], and so on. Two 500MHz niobium cavities have been fabricated by the Shanghai Key Lab of Cryogenics & Superconducting RF Technology at SINAP. The BCP treatment was applied in our cavities as the primary procedure mainly due to its widely understood process technique and lower cost. Besides, HPR has been also used as the final step in cavity preparation to reduce field emission[7]. With surface preparation on #SCD-02 niobium cavity, accelerating gradient at T =4.2K reached as high as 10 MV/m while quality factor is still higher than $4\times10^8$.

## 2 SURFACE PREPARATION SYSTEM

The cavity surface preparation system mainly including a closed-loop BCP and ultra-pure water HPR has been developed at SINAP. The layout of BCP, as shown in Figure 1, mainly consists of acid circulating loop, cooling-water circulating loop, low pressure water rinsing pipeline and recovery pipeline of waste acid, which the most important is that operators of the chemical process were secured against exposure to acids and noxious fumes. Firstly, the acid solution was premixed and cooled by iced flowing water pipe inside acid tank to control the temperature range of 10-15℃. Secondly, the acid pump was started up to circulate the acid through a filter into the cavity and then back to the acid tank. The acid filling time to the cavity corresponding to acid pumping speed was optimized to avoid noticeable shape distortion of lower part of the cavity due to longer exposure to acid. The temperatures of the acid solution and cavity itself were both controlled to achieve a stable etching speed. An iced water shower to the outer cavity surface was used for cooling the cavity. Then, the acid pump was switched off and acid was dumped from the cavity, when the surface layer deep enough was removed. Finally the low pressure water rinsing was immediately carried out to take out the residual acid on inner surface.

---

*Corresponding author (email: liujianfei@sinap.ac.cn)

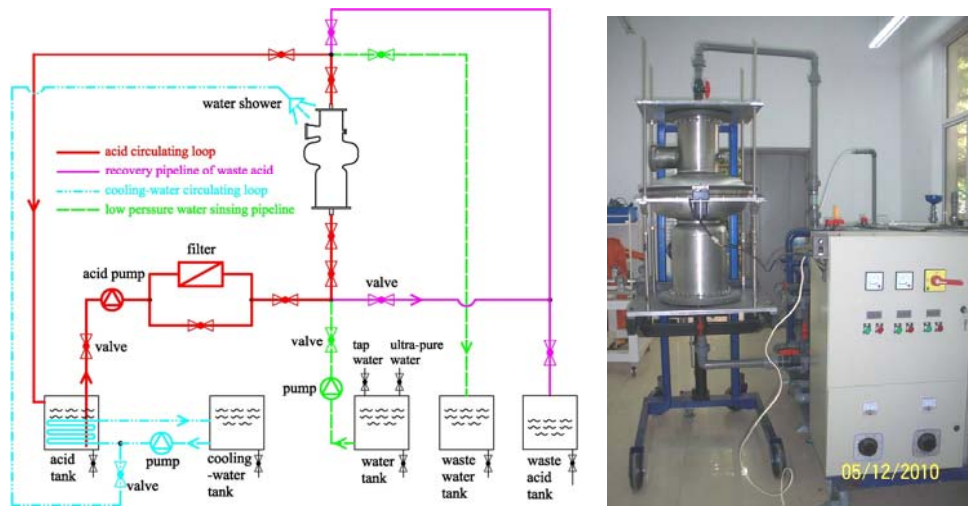

Figure 1. Layout of chemical etching (left) and apparatus (right) at SINAP

A lower etching speed at the cavity equator in vertical BCP came from that a barrier of etched niobium and bubble buffered the further reaction between acid and niobium. A magnetic agitator was added around the cavity equator, which makes use of a rotary turnplate containing magnets outside the cavity to drive the PTFE covered magnetic blocks circling along with the inner wall nearby the equator. The barrier was disturbed and destroyed, so that the etching rate at the equator section was consequently improved. The cavity wall thickness distribution was measured with an ultrasonic thickness meter as shown in Figure 2. It can be seen that the removal thickness was obviously increased at upper equator section where the magnetic agitator was put.

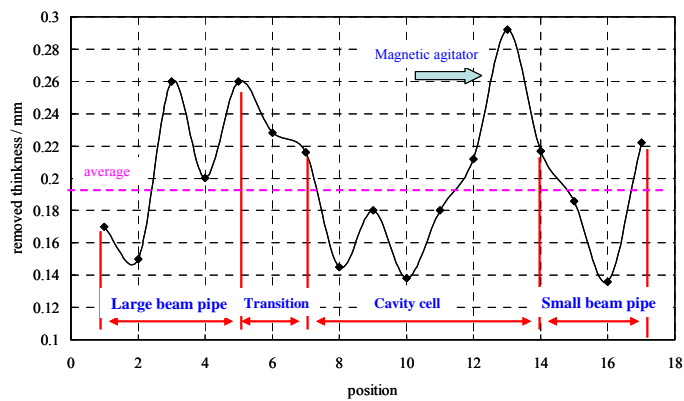

Figure 2. The material removal distribution along the cavity axis with an average of about 190μm

To remove thoroughly chemical residues or particles against field emission, the automated HPR system, consisting of a high-pressure pump, a spray wand, and custom spray nozzles, was applied as the final step in the surface preparation of SC niobium cavities. The ultra-pure water in resistivity of 18 MΩ-cm and pressure of 80kg/cm$^2$ was rinsing from 6 nozzles on top of the feeding cane, which moved up and down inside the cavity while the cavity was rotating.

## 4 CAVITY RESULTS

A typical surface processing procedure was applied to our cavities. BCP was initially performed using a proof-tested 1:1:1.5 solutions of 40% hydrofluoric acid, 65% nitric acid and 85% phosphoric acids premixed in 140 liters. An average total of 190μm of niobium was removed from the inner surface of the cavity during the first BCP treatment of 1.5 hours. This was followed by a 1 hour low-pressure water rinsing. The cavity was dismounted from the BCP apparatus and moved into a class-100 cleanroom for 1.5 hours HPR. Then, drying in a class-10 clean room overnight, a low-power input coupler and a pickup antenna were mounted. The cavity was sealed with vacuum flanges and evacuated. Subsequently the cavity was applied "in-situ bakeout" process at 100℃ for 70 hours. Figure 3 shows the polished inner surface in comparing with that before BCP.

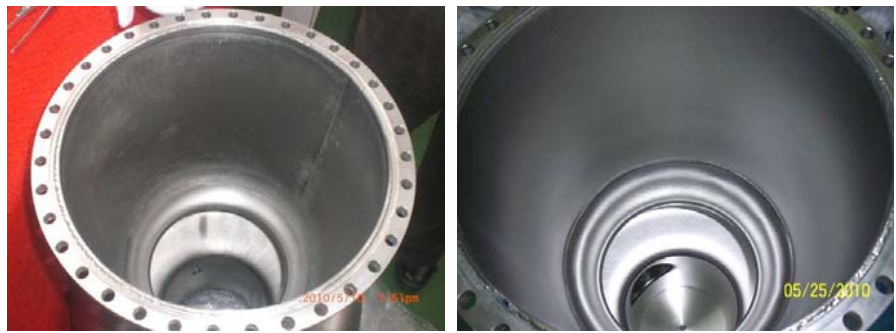

Fgiure 3. The surface comparison before and after heavy BCP+HPR

The first vertical tests were carried out in a vertical cryostat filled with liquid helium at 4.2K, and #SCD-02 cavity showed a strong degradation in quality factor around 4MV/m, as can be seen in Figure 4 (■June 23; ◆June 21).

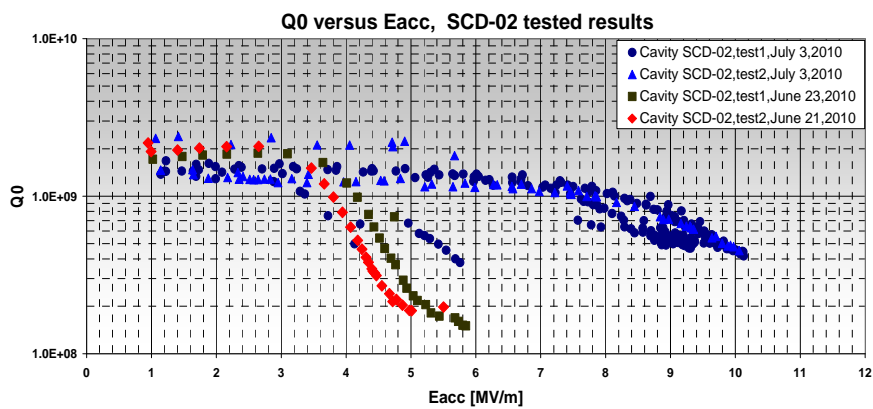

Figure 4. First vertical test results of 500MHz single cell superconducting cavity and the improvements of cavity performance by additional slight BCP

When the cavity was taken off the cryostat and opened to have an inspection, we found the interior of cavity bad smell. It reminded us that there must be a serious contamination during BCP, HPR or cavity assembling, which caused the strong $Q$-drop. The cavity $Q$-drop was cured by improved processing sequence. In July 2, 2010, the second etching process (slight BCP) adopted a new 130 liters acid mixture containing 1 part HF (40%), 1 part $HNO_3$ (65%) and 2 parts $H_3PO_4$ (85%) by volume, and it just took 22 minutes resulting in about 20μm material been removed. After BCP, low-pressure water rinsing had been running for about 80minutes. Subsequently, the wet cavity was taken into class-100 clean room for HPR and assembling. Field gradients of *Eacc* above 10MV/m with quality factors *Qo* above $4\times10^8$ was achieved with second vertical testing in high reliability as shown in Figure 4 (●July 3; ▲July 3).

## 5 Conclusions

The surface preparation is one of the critical steps of SC niobium cavity fabrication. A closed-loop BCP apparatus and an ultra-pure water HPR system had been constructed at SINAP, and been successfully applied to 500MHz superconducting cavities. The 500MHz SC niobium cavities had a simple buffing on the inner cavity surface prior, especially at the welding seam, then went through a sequence of surface processing, including BCP which took about 210μm material removal in total, and effective HPR with fresh ultra-pure water, and low temperature baking at 100℃, the SCD-2# cavity had exceed gradient of 10MV/m with a quality factor $Q0 \geq 4\times10^8$ at 4.2K, satisfying the specification of SSRF superconducting cavities.

## Acknowledgements

*The authors would like to thank Professors Zhao Zhentang and Dai Zhimin from SSRF for their persistent supports on developing the superconducting cavity. The members of the SSRF RF Group are acknowledged for their assistance during the surface preparation processing of 500MHz SC cavities.*

## References

1. F. Furuta, K. Saito, T. Saeki, et al., Experimental comparison at KEK of high gradient performance of different single cell superconducting cavity designs, Proceedings of EPAC, Edinburgh, Scotland, 2006, MOPLS084
2. J. Knobloch., Field Emission and Thermal Breakdown in Superconducting Niobium Cavities for Accelerators, Applied Superconductivity, IEEE Transactions on, 1999, 2: 1016-1022
3. T. Higuchi et al., Investigation on Barrel Polishing for Superconducting Niobium Cavities, Proc. of the 7th workshop on RF superconductivity, Saclay, France, 1995, 723-727
4. M. L. Kinter, I. Weissman, W. W. Stein, Chemical polish for Niobium Microwave Structures, Journal of Applied Physics, 1970, 41: 828
5. T. Higuchi, Development of Horizontal Chemical Polishing for superconducting niobium cavities, 9th Workshop on RF Superconductivity, 1999, Santa Fe, New Mexico USA, tup025
6. K. Saito, et al., Superiority of Electropolishing over Chemical Polishing on High Gradients, Proceedings of the 8th Workshop on RF Superconductivity, Abano Terme, 1997, 759-813
7. J. Knobloch and R. Freyman, Effect of high-pressure rinsing on niobium, Technical report, Cornell University, Laboratory of Nuclear Studies, 1998, SRF report 980223-01
8. P.Kneisel et al., Experience with High Pressure Ultrapure Water Rinsing of Niobium Cavities, Proc. of the 6th Workshop on RF Superconductivity, CEBAF,1993, 628
9. G. Ciovati, Effect of low-temperature baking on the radio-frequency properties of niobium superconducting cavities for particle accelerators, Journal of Applied Physics, 2004, 96: 1591-1600